# THE COOPERATIVE SORTING STRATEGY FOR CONNECTED AND AUTOMATED VEHICLE PLATOONS

Submitted by

**Jiaming Wu**, Postdoc Researcher

Department of Electrical Engineering, Department of Architecture and Civil Engineering,

5433 EDIT Building, Chalmers University of Technology

Gothenburg, SE-41296, Sweden

Email: jiaming.wu@chalmers.se

**Soyoung Ahn**, Professor, **Corresponding Author**

Department of Civil and Environmental Engineering, University of Wisconsin, Madison

2205 Engineering Hall, 1415 Engineering Drive

Madison, WI 53706, United States

Email: sue.ahn@wisc.edu

**Yang Zhou**, Postdoc Researcher

Department of Civil and Environmental Engineering, University of Wisconsin, Madison

2205 Engineering Hall, 1415 Engineering Drive

Madison, WI 53706, United States

Email: zhou295@wisc.edu

**Pan Liu**, Ph.D., Professor

Jiangsu Key Laboratory of Urban ITS, Southeast University

Jiangsu Province Collaborative Innovation Center of Modern Urban Traffic Technologies

Si Pai Lou #2, Nanjing, China, 210096

Email: liupan@seu.edu.cn

**Xiaobo Qu**, Professor

Department of Architecture and Civil Engineering, Chalmers University of Technology

Gothenburg, SE-41296, Sweden

Email: xiaobo@chalmers.se

# The Cooperative Sorting Strategy for Connected and Automated Vehicle Platoons

By Jiaming Wu, Soyoung Ahn, Yang Zhou, Pan Liu and Xiaobo Qu

## ABSTRACT

This paper presents a "cooperative vehicle sorting" strategy that seeks to optimally sort connected and automated vehicles (CAVs) in a multi-lane platoon to reach an ideally organized platoon. In the proposed method, a CAV platoon is firstly discretized into a grid system, where a CAV moves from one cell to another in discrete time-space domain. Then, the cooperative sorting problem is modeled as a path-finding problem in the graphic domain. The problem is solved by the deterministic A* algorithm with a stepwise strategy, where only one vehicle can move within a movement step. The resultant shortest path is further optimized with an integer linear programming algorithm to minimize the sorting time by allowing multiple movements within a step. To improve the algorithm running time and address multiple shortest paths, a distributed stochastic A* algorithm (DSA*) is developed by introducing random disturbances to the edge costs to break uniform paths (with equal path cost). Numerical experiments are conducted to demonstrate the effectiveness of the proposed DSA* method. The results report shorter sorting time and significantly improved algorithm running time due to the use of DSA*. In addition, we find that the optimization performance can be further improved by increasing the number of processes in the distributed computing system.

**Keywords:** Connected and automated vehicles; Vehicle sorting; A* algorithm

# 1. INTRODUCTION

The connected and automated vehicle (CAV) technology is fast developing and demonstrating potential to drastically improve traffic efficiency, stability, and safety. Vehicle-to-vehicle (V2V) and vehicle-to-infrastructure (V2I) *connectivity* can facilitate cooperative system operations based on shared information such as the speed and position of individual vehicles, and traffic signal timing ( *Lee and Park, 2012; Guler et al., 2014*). Vehicle *automation* can facilitate an 'optimal' design of CAV trajectories (e.g., to maximize the throughput at a signalized intersection) and precise control of CAVs according to the design albeit with some uncertainty (*González et al., 2016; Chen et al., 2017; Zhou et al., 2017b; Yang et al., 2018; Chen et al., 2019; Guo et al., 2019*). Numerous studies on CAV control have emerged in the past couple of decades, generally falling into three major areas: longitudinal control, lateral control, and route planning.

**Longitudinal control** primarily focuses on car-following control in CAV platoons to achieve more efficient and stable traffic flow. This type of systems are known as Adaptive Cruise Control (ACC) systems without communication function (*Kesting et al., 2008; Zhou et al., 2017a; Zhou and*

*Ahn, 2019*) or Cooperative Adaptive Cruise Control (CACC) systems with communication (*Milanés et al., 2013; Öncü et al., 2014; Gong et al., 2016; Gong and Du, 2018; Wang et al., 2019a*). The simplest and perhaps most well-known modeling approach is linear feedback control, where a CAV is controlled to maintain a pre-defined constant time headway (*Darbha and Rajagopal, 1999; Rajamani and Shladover, 2001*) or a constant space (*Hasebe et al., 2003*). More recent studies have adopted the optimal control approach, such as model predictive control (MPC), to enable more explicit formulation of an objective function and constraints (*Wang et al., 2014a; Wang et al., 2014b; Zhou et al., 2017a*). This approach can incorporate several objectives, such as control efficiency and driving comfort, into the objective function and directly include critical safety constraints such as collision avoidance.

At a more tactical level, longitudinal control involves a cooperative system between CAVs and traffic infrastructure (*Letter and Elefteriadou, 2014; Talebpour and Mahmassani, 2016; Ma et al., 2017a; Levin et al., 2017a; Li and Zhou, 2017; Zhou et al., 2017b*). For example, *Sun et al. (2017)* proposed a scheme to separate CAV platoons by "movement objectives" at a signalized intersection to maximize the intersection capacity. Rakha and Kamalanathsharma *(2011)* introduced an eco-driving framework, aiming to improve fuel consumption by providing signal control information to CAVs.

**Lateral control** pertains to lane-changing control. A classical method is trajectory planning to generate smooth or "optimal" lane-changing trajectories (*Hatipoglu et al., 2003; Xu et al., 2012*). However, most models of this kind do not consider the reaction of surrounding vehicles, leading to possible conflicts in reality. Some exceptions are notable. For example, Wang et al., *(2015)* addressed potential conflicts by incorporating "proximity costs" in the MPC framework developed in their previous studies (Wang et al., 2014a; Wang et al., 2014b; Zhou et al., 2017c). Luo et al. *(2016)* proposed conflict-free maneuver strategies based on V2V communication. Yang et al., *(2018)* proposed a dynamic trajectory planning model with collision avoidance algorithms. Other existing studies on the lateral control of CAVs are nicely reviewed in Xu et al., *(2012)* and Bevly et al. *(2016)*.

At a more macroscopic scale, **Route planning** aims to find the optimal path for an origin-destination pair. This typically entails solving for the shortest path for the ego-vehicle from its origin to destination using algorithms such as the Dijkstra and A* algorithms (*Fu, 2001; Bell, 2009; Silver, 2005*). With V2V or V2I connectivity, a route can be better planned for the user as well as system optimality (*Du et al., 2015; Melson et al., 2018; Wang et al., 2019b*). Besides this classical pathfinding problem for a single vehicle, the topic of pickup and delivery services with CAVs has emerged recently (*Fleischmann et al., 2004; Levin, 2017b*). For example, Mahmoudi and Zhou *(2016)* proposed an algorithm for solving the pickup and delivery problem with time windows. The optimization problem was formulated into a state-space-time network representation and solved with dynamic programming techniques. Ma et al. *(2017b)* developed a sharing and reservation system with CAVs, where a linear programming model is used for solving the optimal route planning problem. Case studies indicate that the proposed system could significantly increase the vehicle use rate and consequently reduce vehicle ownership.

While there have been promising developments in CAV-related research, some limitations and gaps are notable in the literature. Although longitudinal and lateral control models serve as essential building blocks, they have largely been studied as separate problems. In reality, however, longitudinal and lateral movements often interact, and such interactions should be considered in the same control framework. Although some exceptions exist (e.g., *Contet et al. (2006)*, *Milanés and Shladover (2013)*, and *Bang and Ahn (2018)*), a more systematic integration is necessary for more realistic control. Furthermore, the existing lateral control strategies lack cooperation between a lane-changing vehicle and surrounding vehicles, and among multiple lane-changing vehicles. Similarly, the route planning schemes focus on the navigation problem in the macroscopic scope and lack cooperation for local platoons.

What is largely missing in the literature is strategic "sorting" of CAVs on a multi-lane facility. Indeed, few studies such as Sun et al. (2017) show that capacity of a signalized intersection can be significantly improved if vehicles are organized in a certain fashion according to their anticipated movements (e.g., through, left, right). With the CAV technology, strategic vehicle sorting would be attainable, and the present study explores this opportunity. Our study is further motivated by innovative intersection design concepts such as the tandem design in Fig. 1(a) and the contraflow left-turn lane design in Fig. 1(b). The tandem design uses pre-signals to pre-sort left-turn and through vehicles so that all lanes would be utilized for discharging (thus higher throughput). In the contraflow left-turn lane design, the exit lane is dynamically used as an additional discharge lane for left-turn vehicles with the help of a pre-signal. These designs have gained popularity in certain parts of the world for its effectiveness in increasing the intersection capacity but suffer from unstable performance – an inherent issue in managing human-driven vehicles (*Xuan et al., 2011; Wu et al., 2016; Wu et al., 2019*). With highly cooperative and controllable vehicles, strategic vehicle sorting would be possible, even without pre-signals.

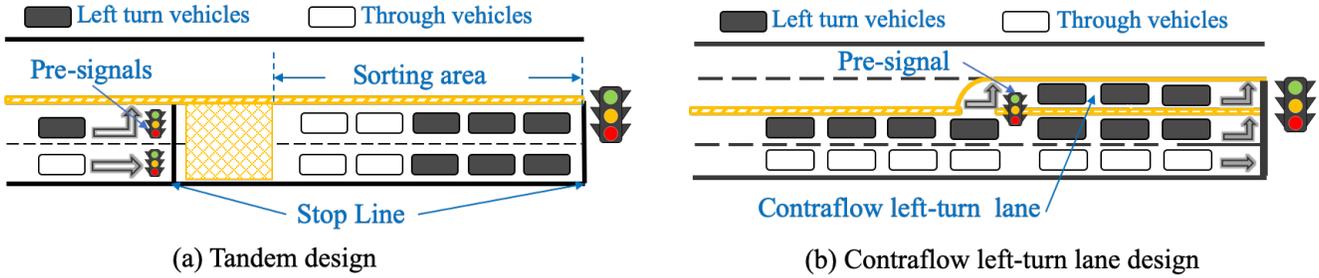

(a) Tandem design         (b) Contraflow left-turn lane design

**Fig. 1.** Practical scenarios for sorted platoons

To this end, the main objective of the present study is to develop a general, yet flexible, vehicle "cooperative sorting" strategy. The proposed strategy aims to optimize the transition of "permutations" – representing relative positions of CAVs in a multi-lane platoon – from any initial permutation to a desired goal permutation. Note that we aim to optimize the sorting process, and thus the solution provides the optimal paths for all CAVs to move to their desired positions while avoiding conflicts. Fig. 2 illustrates the technical scheme of the present paper. A CAV platoon is firstly discretized into a grid system, where a CAV moves from one cell to another in discrete time-space domain. Then the sorting problem is converted to a graphical network, where each node represents a permutation and each edge represents the "cost" associated with all vehicle movements to reach the permutation. Then, it is modeled as a path-finding problem in the graphic domain, which is solved by the deterministic A* algorithm by first assuming that only one vehicle can move at any time step. The resultant shortest path is further optimized with an integer linear programming (ILP) algorithm to minimize the sorting time by allowing multiple movements within a time step. To further improve the algorithm running time and address multiple shortest paths, we also develop a distributed stochastic A* (DSA*) algorithm by introducing random disturbances to the edge costs to break uniform paths (with equal path cost).

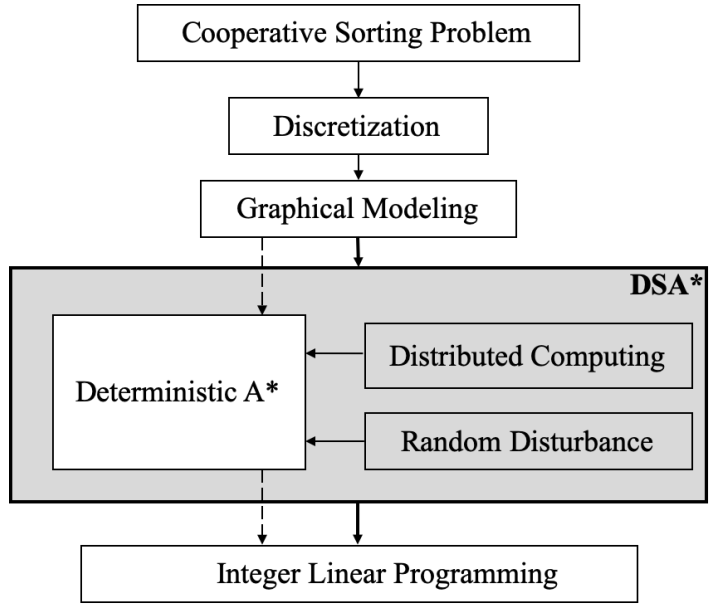

**Fig. 2.** The Technical scheme of the present paper

The rest of the paper is organized as follows. Section 2 describes the basic set-up of the sorting problem and related definitions. Section 3 explains the model formulation including the sorting cost for the sorting optimization. Section 4 further addresses sorting time optimization with the ILP algorithm using the coordinate system designed in this study. Section 5 describes the framework of the proposed DSA* algorithm to efficiently identify multiple shortest paths. A numerical case study is provided in Section 6 to demonstrate the proposed sorting strategy. Finally, the last section concludes the paper with discussions and future research directions.

## 2. THE COOPERATIVE SORTING PROBLEM

This section describes the basic set-up of the cooperative sorting problem considered in this paper. Detailed formulation and solving algorithms will be presented later. The main problem we address is to sort randomly distributed CAVs in a platoon to achieve an ideally-organized platoon. Without loss of generality, we will focus on an isolated road segment that can be connected to any downstream facilities, e.g., signalized intersections or freeways. The basic set-up is as follows: (a) a platoon consists of CAVs clustered across multiple lanes; (b) the study period begins with a random arrival of a platoon and ends with an ideally-sorted platoon; (c) the number of vehicles in the platoon is constant during the analysis, and any vehicle joining or leaving the platoon will lead to a re-start of the proposed algorithm; and (d) in the initial platoon, the vehicular speed could be different among CAVs, but the sorted platoon will have a uniform speed, named the "cruising speed" for simplicity. Note for (a) that we assume 100% CAVs for simplicity, but connectivity alone (without automation) would be sufficient as long as human-driven vehicles comply. We also understand that 100% CAVs will not be seen in the near future, but the proposed sorting strategy could be a building block to deal with mixed traffic conditions, which is part of our ongoing research.

System state, $S_i$, represents a permutation of CAVs in a platoon, which changes as vehicles change their relative positions in the platoon. The cost of transition from state $i$ to state $j$ for vehicle $n$ is denoted as $C_{ij}^n$. This cost is determined based on the "effort" (e.g., travel distance, travel time, or consumed fuel) that vehicle $n$ has to make to change its position according to the state change. Then, the objective of vehicle sorting can be formulated as

$$\text{Min } \sum_{i=1}^{N} C_{IG}^{i} \qquad (1)$$

subject to

$$C_{IG}^{i} = f_i(S_G | S_I) \qquad (2)$$

$$f_i(S_G | S_I) \in \text{U} \qquad (3)$$

where $I$ denotes the initial state; $G$ denotes the desired goal state; $N$ denotes the total number of vehicles in the platoon; $f_i(S_G | S_I)$ is the specified cost function; and $U$ is the kinematical constraints. The tandem concept in Fig. 1(a) will be used as an example in this paper to illustrate the proposed method. In the example, left-turn and through vehicles need to be separated into two groups, either the left-turn vehicles in front of the through vehicles or the opposite. The sorting process should be completed just upstream of the intersection, as shown in Fig. 3. However, unlike the conventional tandem design, vehicle sorting will be achieved without pre-signals.

In Fig. 3, a goal state corresponds to a permutation where all left-turn vehicles queue in front of through vehicles, anticipating leading left-turn signal. As lateral positions within a group need not be specific, there exist a combinatorial number of permutations that satisfy the tandem design, leading to multiple goal states. This issue will be addressed in Section 3.2.2. Note that the proposed strategy deals with path planning of how CAVs should change their relative positions in the platoon. Detailed trajectory planning to execute the sorting process is beyond the scope of the present paper.

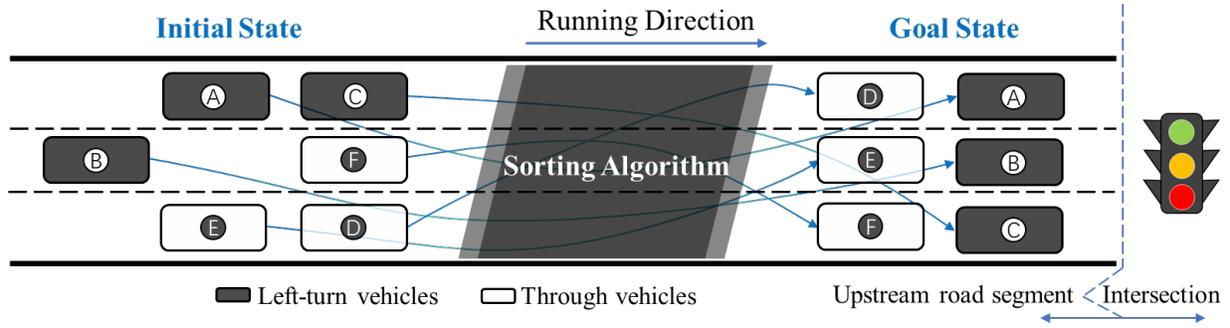

**Fig. 3.** The cooperative sorting problem with the tandem design concept

# 3. MODEL FORMULATION FOR THE SORTING COST OPTIMIZATION

In this section, we formulate a stepwise algorithm for solving the cooperative sorting problem while considering vehicle kinematical constraints. To this end, we firstly discretize the problem in the time-space domain by defining the system state (vehicle permutation) in a grid system and updating the state at discrete time steps. Cooperative sorting is then modeled as a path-finding problem in the graphic domain and solved by a stepwise deterministic A* algorithm. Note that since we focus on the vehicle position relative to other vehicles, not to a fixed point on a road segment, we adopt a moving view relative to the cruising speed for convenience of analysis.

## 3.1 Discretization and Graphic Modeling

A road segment is divided into homogeneous cells in a grid system according to the following guidelines:

(a)     Each lane is represented as a column in the grid system.

(b)     The cell width is equal to the lane width.

(c)     The cell length is set up as the critical gap required to complete a lane-changing maneuver when

all vehicles travel at the cruising speed. The desired gap would vary in practice, depending on the vehicle's mechanical dynamics, speed, and position, etc. However, for the purpose of developing a modeling framework, we assume that the cell length equals a constant value, $L$.

(d)    At most one vehicle can occupy a cell. This is inherently ensured by defining the cell length as in (c). Existing studies indicate that the critical gap required for lane-changing maneuvers by human drivers is very short when surrounding vehicles have similar speed (*Hidas, 2002; Toledo et al., 2003*). Since we consider CAVs that drive cooperatively, the critical gap could be even shorter.

(e)    Each vehicle is assigned to only one cell according to the position of the vehicle midpoint. Note that the exact vehicle position inside the cell is not the focus of the present study and therefore, is not specifically addressed. Furthermore, vehicles may not perfectly line up laterally in practice, but we assume perfect alignment for the sake of modeling.

(f)    Each vehicle can only move into an adjacent cell that is vacant or remain in the same cell.

(g)    Only vertical and horizontal movements are allowed, where the former represents a change in the longitudinal position in the same lane, and the latter represents a lane-changing maneuver. Note that lane-changing can be represented by either a horizontal movement or a diagonal movement, depending on the problem setup. Since we adopt a moving view relative to the cruising speed, only horizontal movement is used to represent lane-changing for consistency.

The size of the grid system (in terms of the total number of cells) should be large enough to ensure sufficient vacant cells for vehicle maneuvers. Ideally, the number of vacant cells should be greater than or equal to the total number of vehicles. This can be achieved by including additional rows of empty cells upstream or downstream of the platoon. For the example in Fig. 3, four rows and three columns will suffice. With assumptions (a)-(g), the positions in the platoon are simplified as shown in Fig. 4. The discretized grid system evolves by "movement step," defined as a position change when a vehicle leaves its current cell. Therefore, the grid system will be static if all CAVs travel at the cruising speed and remain in their current cells.

For modeling, the discretized two-dimensional grid system can be formulated as a matrix, **A**, called the "permutation matrix" hereafter. In the permutation matrix, vacant cells are denoted with 0 and occupied cells are denoted with vehicle IDs. The formulation of the permutation matrix enables calculation of the cost between any states, which will be described in detail in the following section. The sorting process and the corresponding state evolution can then be formulated as matrix transition as illustrated in Fig. 5.

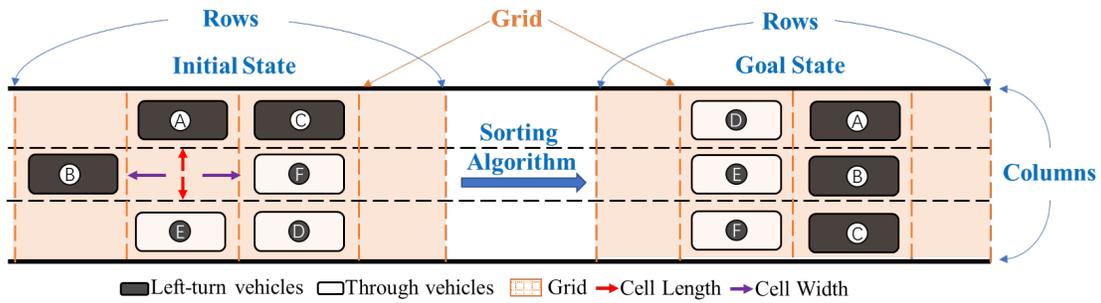

**Fig. 4.** Discretization of the CAV platoon into a grid system

In the matrix scheme, the state space, **S**, can be created containing all possible permutation matrices. In the state space, if state A can be achieved within one movement step from state B, state A and state B are defined as adjacent states. Clearly, there is only one unique movement that links two adjacent states. Thus, the state space can be modeled as a graph system.

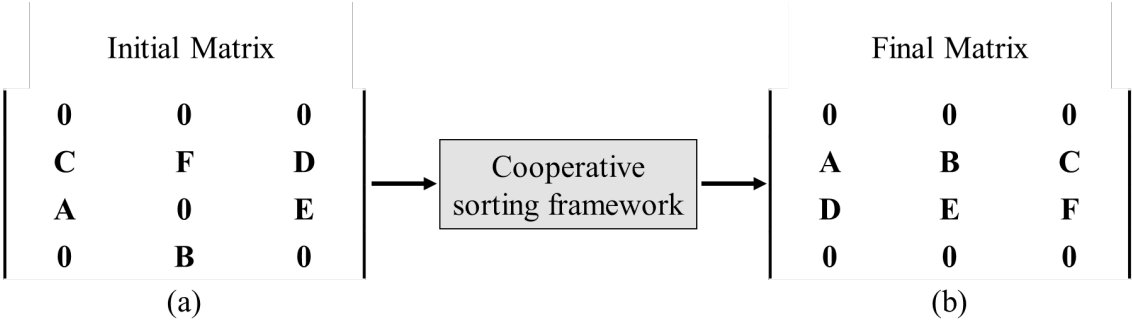

**Fig. 5.** The cooperative platoon sorting problem: transition of the permutation matrix

A basic graph system is usually defined as $G\ (V, E)$, where $V$ denotes nodes and $E$ denotes edges linking these nodes. In cooperative sorting, each permutation matrix $\boldsymbol{A}$ is considered as a node, and the unique movement linking adjacent states $i$ and $j$ is considered as an edge with cost $C_{ij}$. Therefore, state space $\boldsymbol{S}$ can be modeled as a graph system, $\boldsymbol{S}\ (A, C)$. Then, the cooperative sorting problem can be solved by finding the shortest path that links the initial state to the goal state in $\boldsymbol{S}\ (A, C)$. This representation is illustrated in Figure 6.

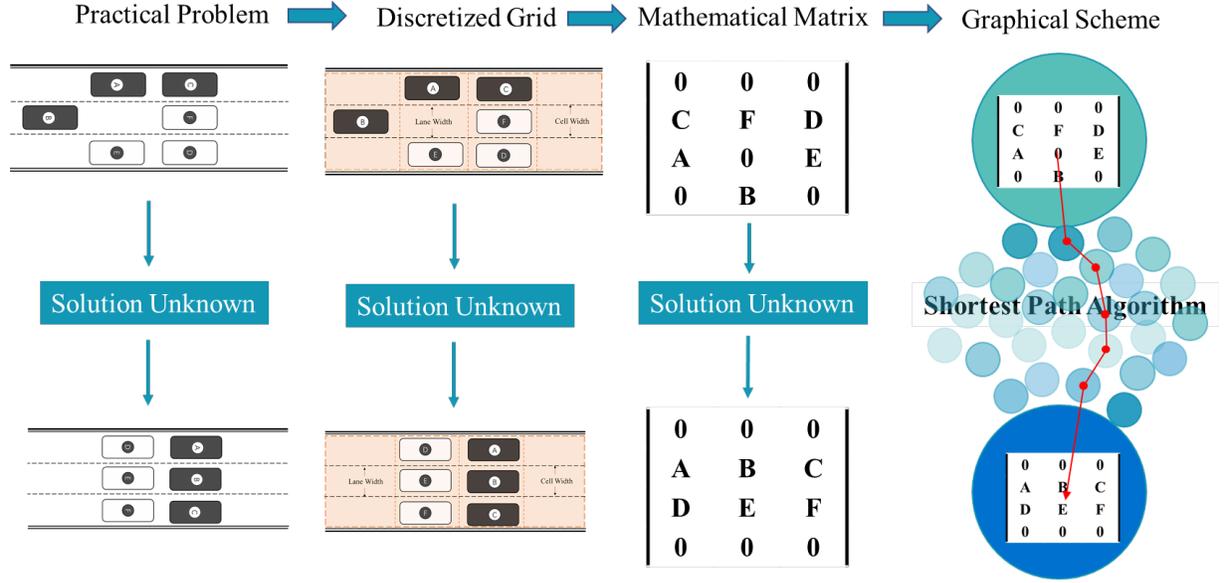

**Fig. 6.** The shortest path modeling procedure

In the graph representation, the edge cost, $C_{ij}$, equals the total efforts that all vehicles make for holding or changing their current positions. For any vehicle, there are three possible maneuvers: holding its relative position, making a longitudinal movement, and changing lanes. Assuming that all CAVs tend to approach the cruising speed, the cost for one vehicle changing from one state to another can be specified as:

$$f_i(S_{K+1}|S_K) = w_h^i + w_l^i + w_{LC}^i \tag{4}$$

$$w_h^i = \beta_{po}\big(\exp(\gamma|v_{i,k}|) - 1\big) \tag{5}$$

$$w_l^i = (1 - \beta_{po})\{\beta_{long} + |v_{i,k}|\exp[v_{i,k}(x_k^i[1] - x_{k+1}^i[1])]\} \tag{6}$$

$$w_{LC}^i = \beta_{lc}f_{lc} \tag{7}$$

Where

$$v_{i,k+1} = \begin{cases} 0 & if\ v_{i,k}^2/2L \leq |a_{limit}| \\ \pm\sqrt{v_{i,k}^2 + 2a_{max}L} & if\ v_{i,k}^2/2L > |a_{limit}| \end{cases}, \quad a_{limit} = \begin{cases} a_{max}^a & if\ v_{i,k} \geq 0 \\ a_{min}^d & if\ v_{i,k} < 0 \end{cases} \quad (8)$$

$$f_{lc} = \begin{cases} 1 & if\ x_k^i[2] \neq x_{k+1}^i[2] \\ 0 & if\ x_k^i[2] = x_{k+1}^i[2] \end{cases}, \beta_{po} = \begin{cases} 1 & if\ x_k^i[1] \neq x_{k+1}^i[1] \\ 0 & if\ x_k^i[1] = x_{k+1}^i[1] \end{cases} \quad (9)$$

$$0 \leq \beta_{long} \leq 1, 0 \leq \beta_{lc} \leq 1, \gamma > 0 \quad (10)$$

$w_h^i$, $w_l^i$ and $w_{LC}^i$ are the costs of vehicle $i$ holding the current longitudinal position, moving in a longitudinal direction and making a lane-change, respectively; $L$ is the cell length (*m*); $v_{i,k}$ is the relative longitudinal speed of vehicle $i$ to the cruising speed in state $S_k$ (*m/s*); $a_{limit}$ is the maximum acceleration, $a_{max}^a$, or minimum deceleration rate, $a_{min}^d$, (*m/s²*); $x_k^i[1]$ and $x_k^i[2]$ represent the longitudinal and lateral positions in the grid matrix of vehicle $i$ in state $S_k$, respectively; and $\gamma$, $\beta_{po}$, $\beta_{long}$, and $\beta_{lc}$ are weight coefficients. Note that the cost function should depend on the objectives. Here, the cost function is designed considering the efficiency of maneuvers and vehicular dynamic constraints.

**Remark 1**: Eq. (4) is designed to quantify the cost of various maneuvers and avoid unlikely movements, such as: (i) maintaining the vehicle's longitudinal position with a large relative speed and (ii) moving in an opposite direction to its large relative speed (e.g., moving backwards in the relative space with speed greater than the cruising speed). For (i), the first term of Eq. (4), $w_h^i$, will generate large costs for all edges connected to this node, so that a path-finding algorithm will exclude this node in the solution. Similarly for case (ii), the second term in Eq. (4) will assign large costs to the edges connected to the node, effectively eliminating the node in the solution.

## 3.2 The Deterministic A* Algorithm

Shortest path finding is a classic problem in graph theory that could even be solved manually for simple graphs. However, the state space, *S (A, C)*, is huge for our study: for the example in Fig. 3, there are 665,280 nodes (representing different platoon permutations). Thus, an efficient solving algorithm is necessary. In our study, we adopt the A* algorithm by Hart et al. *(1968)* for its flexibility and efficiency. The basic idea of the algorithm is to find the shortest path based on an estimated distance to the objective. The estimated distance is usually determined with some background knowledge to the problem (e.g., the sum of distances from the current to the goal positions for all vehicles) to provide some direction where the desirable path might be. The A* algorithm then only needs to expand the subgraph directed by the estimated distance rather than examining the entire graph. Therefore, the A* algorithm is usually more efficient than other algorithms such as classical Dijkstra's method.

A general form of the estimated distance is defined with a cost function, consisting of the real distance and the heuristic distance:

$$\hat{F}(x) = \hat{G}(x) + \hat{H}(x) \tag{11}$$

where $\hat{F}(x)$ is the cost function for the estimated distance to the goal node if the path goes through node $x$; $\hat{G}(x)$ is the real distance function, denoting the shortest distance so far from initial node $I$ to node $x$; $\hat{H}(x)$ is the heuristic distance function, denoting the estimated distance from node $x$ to the goal node. The details of $F(x)$ will be discussed in section 3.2.1. First, we introduce the basic steps of the algorithm as follows:

(1) Create an open list and put the initial state $A_I$ into the open list;

(2) Calculate the *F* values for all nodes in the open list and select the one with the smallest *F* value as the current node.

(3) If the current node is the goal node, terminate the algorithm and return the shortest path.

(4) Otherwise, put the current node to a closed list and put its adjacent nodes that are not already in the closed list into the open list. Then repeat step 2.

An example is provided in ***Appendix A*** to illustrate the procedure of the A* algorithm. In the cooperative sorting problem, there might be multiple ways to expand adjacent nodes. In this paper, if several adjacent nodes are expanded from node *x*, we refer to node *x* as the *parent node* and generated nodes as the *children nodes*. When all *N* vehicles are allowed to move in each movement step, there will be $5^N$ possible children nodes for each parent node, and thus, the complexity of the algorithm grows exponentially with the number of vehicles. For better computational efficiency, we will initially adopt the "stepwise-principle," where only one vehicle is allowed to move in each movement step, limiting the number of children nodes to 5. This assumption will be relaxed in section 5 for the practical scenario where vehicles move simultaneously.

It is obvious that the heuristic function, $\hat{H}(x)$, is critical for the performance of the A* algorithm. Some important features of the heuristic function have been proposed and proved in Hart et al. *(1968)*, which are summarized as follows.

***Definition 1***: An algorithm is called *admissible* if it is guaranteed to find the shortest path.

***Theorem 1***: The A* algorithm is *admissible*, if the relationship in Eq. (12) holds, where $H(x)$ is the distance from node $x$ to the goal node $T$ in the shortest path.

$$\widehat{H}(x) \leq H(x) \qquad (12)$$

***Theorem 2***: If $\widehat{H}(x)$ satisfies the triangle inequality as shown in Eq. (13), then the A* algorithm is optimal among all algorithms with the same background information, where *m* is the current node of interest, *n* is another random node in the graph, and *H(m, n)* is the optimal path from *m* to *n*. Such a heuristic function is called a *consistent* heuristic function.

$$H(m,n) + \widehat{H}(n) \geq \widehat{H}(m) \qquad (13)$$

**Theorem 1** guarantees that A* algorithm could always find the solution if the heuristic distance never overestimates the actual distance of the optimal path. **Theorem 2** ensures that if the heuristic distance is consistent, then the A* algorithm would explore the minimum number of nodes among all searching algorithms. In other words, if the estimated distance cannot be improved by introducing a new node, then the background information is optimally used and the A* algorithm is also optimal. Generally, if the heuristic function is defined with a uniform rule for all nodes, the consistency law would be satisfied (*Hart et al., 1968*). We shall now specify the detailed heuristic functions that will be used in the cooperative sorting problem.

**3.2.1 Heuristic functions**

As defined above, an ideal heuristic function should be both admissible and consistent to guarantee that the A* algorithm could find the shortest path with the minimum number of explored nodes. Considering that the cooperative sorting problem is discretized as a grid system, two widely known admissible and consistent heuristic functions for grid systems are considered: the Manhattan distance and the misplaced distance. In the cooperative sorting problem, modified Manhattan and misplaced distances are proposed as follows:

$$\hat{H}_{ma}(K,G) = \sum_{j=k}^{G} \sum_{i=1}^{N} f_i\left(S_{j+1} \middle| S_j, v_{i,j} = 0\right) = \sum_{i=1}^{N}\left[\left|x_G^i[1] - x_K^i[1]\right|\beta_{long} + \left|x_G^i[2] - x_K^i[2]\right|\beta_{lc}\right] \quad (14)$$

$$\hat{H}_{mis}(K,G) = nC_{min} = \min\left(\beta_{long}, \beta_{lc}\right)n \quad (15)$$

where $\hat{H}_{ma}(K,G)$ and $\hat{H}_{mis}(K,G)$ denote the Manhattan distance and misplaced distance respectively from state $K$ to the goal state $G$; $n$ denotes the number of vehicles that are not in their goal position; $C_{min}$ denotes the minimum edge cost; and $\beta_{long}$ and $\beta_{lc}$ has the same meaning as defined in Eq. (4) to Eq. (10). The misplaced distance is defined by the number of vehicles that are not in their goal positions. A simple example in Fig. 7 illustrates the calculation of the Manhattan and misplaced distances. In Fig. 7, two vehicles are not in their final position in the goal state. The Manhattan distance can be calculated as $\left(|1-2|\beta_{long} + |2-3|\beta_{lc}\right) + |4-3|\beta_{long} = 2\beta_{long} + \beta_{lc}$, as illustrated in Fig. 7(a). In contrast, the misplaced distance is $2\min\left(\beta_{long}, \beta_{lc}\right)$ as only two vehicles are not in their goal positions; see Fig. 7(b).

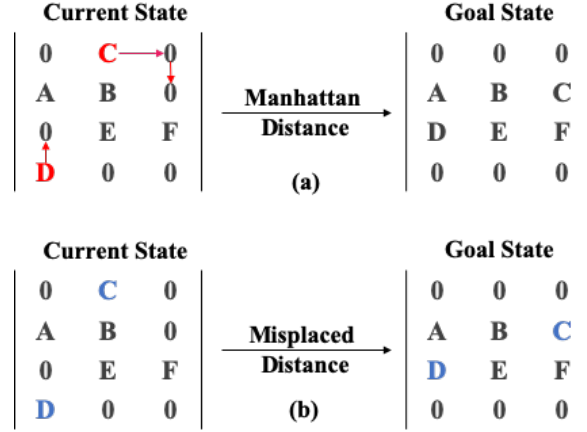

**Fig. 7.** Examples of the heuristic distance in a uniform grid

The misplaced distance defined in Eq. (15) can be considered as a special case of the Manhattan distance, in that each vehicle can move to their goal position with one step. Therefore, we will only prove that the Manhattan distance is admissible and consistent with the following propositions.

***Proposition 1***: The Manhattan distance defined in Eq. (14) is admissible in the cooperative sorting problem.

***Proof***: This proposition is intuitively true because the Manhattan distance is the shortest path for each vehicle. In addition, Eq. (14) assumes an ideal condition that all vehicles travel at the cruising speed, leading to underestimated edge costs, as shown in Eq. (16). Eq. (14) also indicates that all vehicles go through the Manhattan path to their goal positions regardless of trajectory conflicts. In practice, the cooperative sorting process almost always needs some detour, especially with vehicle platoons in high density. Any detours would incur extra costs. Therefore, Eq. (17) can be derived to prove that the proposed Manhattan distance is admissible.

$$f_i\big(S_{j+1}\big|S_j, v_{i,j} = 0\big) \leq f_i\big(S_{j+1}\big|S_j\big) \tag{16}$$

$$\hat{H}_{ma}(K, G) = \sum_{j=k}^{G} \sum_{i=1}^{N} f_i\big(S_{j+1}\big|S_j, v_{i,j} = 0\big) \leq \sum_{j=k}^{G} \sum_{i=1}^{N} f_i\big(S_{j+1}\big|S_j\big) \leq H(K, G) \tag{17}$$

***Proposition 2***：The proposed Manhattan distance is consistent in the cooperative sorting problem.

***Proof***: For one vehicle, the Manhattan distance provides the shortest path to its goal position. Assume that node $n$ is another node in the graph. If all vehicles in the state corresponding to node $n$ are in the Manhattan path of node $m$ to the goal node $G$, then we have:

$$H_{ma}(m, n) + \hat{H}_{ma}(n, G) = \hat{H}_{ma}(m, G) \tag{18}$$

However, if any vehicles are not in the Manhattan path of node $m$, then the following inequality is achieved.

$$H_{ma}(m, n) + \hat{H}_{ma}(n, G) > \hat{H}_{ma}(m, G) \tag{19}$$

Therefore, in all circumstances, we have proved that the Manhattan distance is consistent, as shown below:

$$H_{ma}(m, n) + \hat{H}_{ma}(n, G) \geq \hat{H}_{ma}(m, G) \tag{20}$$

Generally, the proposed Manhattan and the misplaced heuristic functions can be used in almost all conditions in traffic management. However, it should be noted that the Manhattan distance requires a paired relationship between initial and goal states, with a unique position for every vehicle in the initial and goal permutation matrices. In practice, this is not always true. For example, at a signalized intersection, it suffices in the situation in Fig. 7 that vehicles A, B, and C are placed ahead of vehicles

D, E, and F, regardless of their relative positions in the same group. For the cases like this without a clear pair relationship, the misplaced distance should be used. This leads to the insight that the Manhattan distance is preferred when each vehicle has a unique destination or desired position (e.g., if multiple downstream intersections or ramps are considered), whereas the misplaced distance is more suitable for local-level platoon sorting problems.

### 3.2.2 Multiple goal states

In the previous sections, we assumed that there is only one goal state in our cooperative sorting problem. In practice, however, only a general objective may exist, such as minimizing the number of lane-changes or separating platoons between left and through movements. In these cases, it is not always easy to determine a unique goal state, especially when the cost of approaching the unique state is unknown. At signalized intersections, as exemplified in Fig. 8, both conventional and tandem designs could be feasible for signal control plans. In addition, there are three variants for the tandem design because the number of left-turn vehicles cannot be exactly divided by the number of lanes. An ideal solution is that the user selects multiple goal states, and the algorithm returns the optimal goal states with the minimized cost.

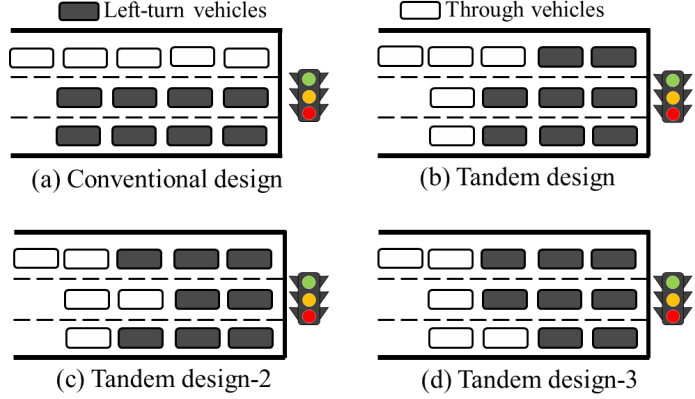

**Fig. 8.** Multiple goal states

An intuitive way of solving this problem is to consider multiple goal states as a goal state set. In the cooperative sorting problem, this idea could be achieved by creating a dummy goal node that is only connected to the goal state set, as shown in Fig. 9. The goal state in the A* algorithm framework can then be set as the dummy node, so that the optimal path would surely go through one of the multiple goal states. Another notable setup is the heuristic function, as in Eq. (21), to be admissible, where $\Theta$ is the goal state set. The dummy node setup also provides more flexibility in the goal state modeling. For instance, if the multiple goal states are equally preferred, then the edge cost from the multiple goals to the dummy goal node can be assumed to be zero. However, if some goal states are preferred to others, weighted coefficients can be setup for different goal states.

$$\hat{H}_{ma}(K, G) = \min_i \hat{H}_{ma}(K, G_i), \, G_i \in \Theta \tag{21}$$

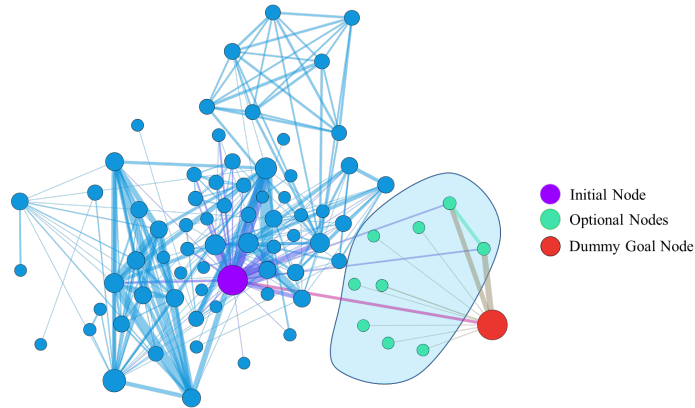

**Fig. 9.** The dummy goal node concept

# 4    SORTING TIME OPTIMIZATION

Besides the total sorting cost, minimizing the sorting time is also important for practical implementation. An intuitive way to achieve this is to allow simultaneous movements of multiple vehicles, named "multi-movement principle" hereafter. Note that optimizing for sorting cost and time simultaneously is not feasible as most path-finding algorithms are cost-based. Therefore, we propose a two-step framework for minimizing the sorting cost and time: (1) solve the problem based on the stepwise-principle, which returns the minimum sorting cost; and (2) optimize the trajectories without increasing the cost with the multi-movement principle, which returns the minimum sorting time.

The basic idea of the sorting time optimization is to reduce unnecessary waiting time in each vehicular movement based on the multi-movement principle. In addition, the setpoints of vehicles' trajectory should be maintained without extra cost while avoiding conflicts. This type of problem can be solved by integer linear programming (ILP), where the objective function is defined to minimize the number of sorting steps, and constraints are formulated to avoid trajectory conflicts.

For mathematical modeling, we design a three-dimensional coordinate system to track the position changes of every vehicle. The coordinate system is illustrated in Fig. 10, and the modeling procedure is provided as follows:

(1) Convert the permutation matrices to row vectors; see Fig. 10 for the example, where a $4 \times 3$ matrix is converted to a vector with the length of 12.

(2) Track the position of each vehicle via the coordinate system, $T_i(x, y) = s$, where $i$ is the vehicle index, $x$ and $y$ respectively represent the original and new positions in the permutation vector, and $s$ is the step at which the position changes. In Fig. 10, vehicle F changed its position in step 1. Therefore, it is denoted $T_F(5,8) = 1$ in the proposed coordinate system.

(3) If the vehicle did not change its position, then $T_i(x, y) = -1$, where $x = y$.

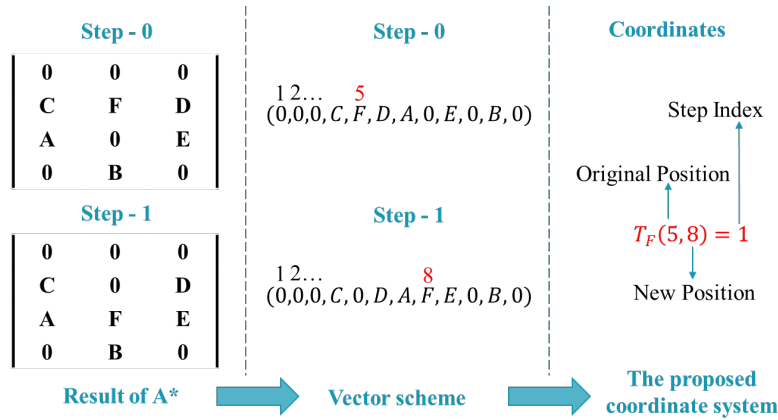

**Fig. 10.** The proposed coordinate system

We consider two levels of the multi-movement principle: "conservative" and "aggressive". In the *conservative* level, vehicles can move into a cell only if it is empty in the last step. The corresponding ILP problem can be formulated as follows:

$$\min \sum_i T_i (m_i, n_i) \tag{22}$$

**Subject to**
$$T_i(m_i, n_i) = \max_i T_i(x, y) \tag{23}$$

$$T_i(x, y) < T_i(y, z) \tag{24}$$

$$T_i(x, y) < T_j(z, x), \forall\, i \neq j \tag{25}$$

$$T_i(x, y) \geq 1 \tag{26}$$

$$x, y, z \in \Phi,\ T_i(x, y) \in \mathbb{Z}^+ \tag{27}$$

where $\Phi$ denotes a set of positions associated with position changes. The objective function in Eq. (22) aims to minimize the sum of step indices when all vehicles arrive at their goal positions. The constraints in Eq. (24) and (25) are formulated to avoid conflicts. Specifically, the constraint in Eq. (24) dictates that the order of the positions that a vehicle occupies is consistent with the result of the A* algorithm. The constraint in Eq. (25) ensures that in each position, the sequence of vehicles that occupy the position is also consistent.

In the *aggressive* level, the ILP problem can be formulated as follows:

$$\min \sum_i T_i(m_i, n_i) \tag{28}$$

**Subject to**
$$T_i(m_i, n_i) = \max_i T_i(x, y) \tag{29}$$

$$T_i(x, y) < T_i(y, z) \tag{30}$$

$$T_i(x, y) \leq T_j(z, x), \forall\, i \neq j \tag{31}$$

$$T_i(x, y) \geq 1 \tag{32}$$

$$x, y, z \in \Phi,\ T_i(x, y) \in \mathbb{Z}^+ \tag{33}$$

The only difference lies in Eq. (31), which allows a vehicle to leave a cell and another vehicle to fill it in the same movement step.

# 5    THE DISTRIBUTED STOCHASTIC A* ALGORITHM

The previous sections provided the basic theoretical framework for solving the cooperative sorting problem. This section addresses computational efficiency of the proposed method for real-time implementation. This is a major challenge due to the extensive combinatorial graph space of the problem and existence of multiple shortest paths. The latter happens when multiple CAVs are driving at the same cruising speed, resulting in the same costs for their longitudinal movements and lateral movements, though the lateral cost would be different from the longitudinal cost. In the graph space, it means that many edges will have the same cost and form uniform subgraphs. Thus, multiple shortest paths exist, and the deterministic A* algorithm has to explore every node with the same estimated cost, which might increase the running time significantly. Another limitation of the deterministic A* algorithm is that it only returns one shortest path based on some fixed rule, such as the minimum node ID, without further ILP optimization. To remedy these shortcomings, we develop a distributed stochastic A* (DSA*) algorithm with a hierarchical and stochastic heuristic function to improve the solving algorithm's performance in terms of running time and optimization. Section 5.1 will introduce the framework of the DSA* algorithm, and Section 5.2 will elaborate on the design of the stochastic heuristic function.

## 5.1 The Distributed Framework

The basic idea of the DSA* is to introduce a random disturbance factor in the heuristic function so that the equal shortest paths in the deterministic A* will have slightly different costs. This treatment will break uniform subgraphs, thereby reducing the number of explored nodes and thus the running time. Furthermore, the algorithm will return multiple equal shortest paths in different iterations. With distributed computing techniques, the algorithm can generate different shortest paths simultaneously as inputs for the ILP optimization (that employs the multi-movement principle).

The distributed system is illustrated in Fig. 11. In the proposed system, the DSA* algorithm is run in a distributed way, where each CAV serves as a computing center, and parallel computing techniques are used within each vehicle. Multiple shortest paths will then be found by the DSA* algorithm and further optimized by the proposed ILP algorithm. Eventually, the shortest path with the minimum steps will be selected as the final solution. In practice, the algorithm efficiency can be further improved by techniques such as the "time limit" setup that can terminate solutions taking too long and start a new search, depending on the urgency. Obviously, a smaller time limit would enhance computational performance, albeit at the expense of optimization performance due to fewer shortest paths that can be considered in the ILP optimization.

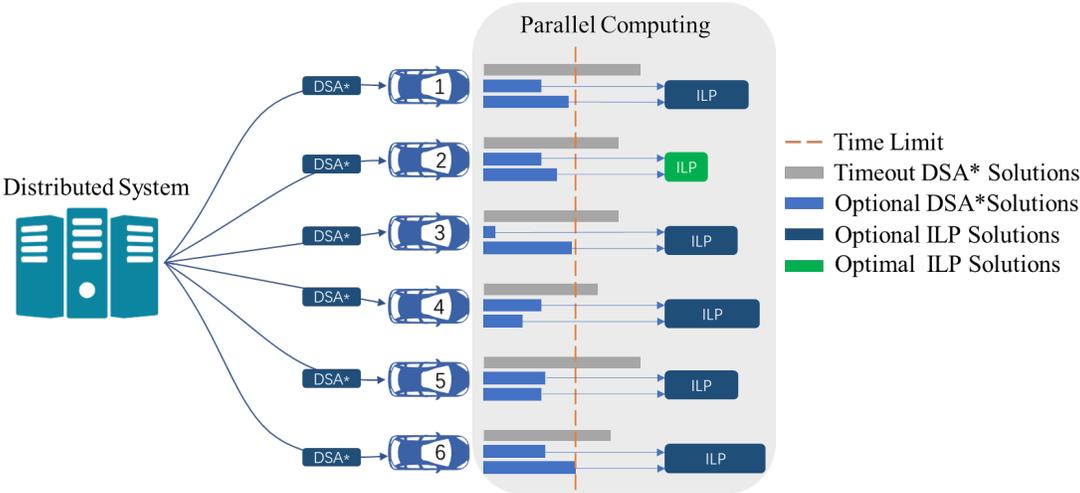

**Fig. 11.** Illustration of the distributed stochastic A* (DSA*) concept

## 5.2    The Hierarchical Stochastic Heuristic Function

For the A* algorithm and its variants, the search performance is mainly decided by the heuristic

function. For the DSA* algorithm, we design a hierarchical stochastic heuristic function as follows:

$$\hat{H}_{DSA*} = \hat{H} - \varepsilon C_{\min}/\exp\left(\hat{H}\right), \varepsilon \in (\exp(-C_{\min}), 1) \tag{34}$$

where $\hat{H}_{DSA*}$ denotes the modified heuristic function; $\hat{H}$ denotes the heuristic function in the deterministic A* algorithm; $\varepsilon$ is a random factor; and $C_{\min}$ is the minimum edge cost. Note that the range of $\varepsilon$ should be a function of the minimum edge cost to maintain optimality while improving the algorithm efficiency. Several important properties are considered in designing this function and proved in the following propositions. For convenience, we consider that the graph consists of different layers according to the distance to the initial node. Therefore, the graph is generally explored from inner layers with small $G$ values to outer layers with large $G$ values. (Recall that $G$ value indicates the total

cost associated with the shortest path found so far from the initial node to the current node.)

**Proposition 3**: The $\hat{H}_{DSA*}$ maintains the same admissibility as $\hat{H}$.

**Proof**: if the deterministic heuristic function $\hat{H}$ is admissible, $\hat{H}_{DSA*}$ is obviously admissible based on

***Theorem 1*** and Eq. (35).

$$\hat{H}_{DSA*} = \hat{H} - \varepsilon C_{\min}/\exp\left(\hat{H}\right) \leq \hat{H} \leq H \tag{35}$$

**Proposition 4**: For ununiform graphs, $\hat{H}_{DSA*}$ does not undermine the efficiency of $\hat{H}$ with the

Manhattan and misplaced distance concepts.

**Proof**: The heuristic function directs a way to search the unexplored parts of the graph. An overestimated heuristic function puts more weights on future paths than explored ones and thus aggressively explore the path to outer layers, which leads to a fast but possibly inadmissible search. On the contrary, an underestimated (admissible) heuristic function ensures optimality but prefers inner layers because expanding the path will increase the $F$ value (estimation of the shortest path), leading to diminished search efficiency. For admissible heuristic functions, the A* algorithm will be more inclined to stay within inner layers and existing nodes with a smaller the heuristic function. Therefore, we need to prove that the modified heuristic function will not further prefer existing nodes, even though it decreases the edge costs by random factors. Assuming that the path moves from current node $i$ to adjacent node $j$, this condition dictates that node $j$ has the minimum $F$ value in the open list $O(i)$, as formulated in Eq. (36).

$$\hat{F}(j) < \hat{F}(k); \hat{F}(j) = \hat{G}(j) + \hat{H}(j); \hat{F}(k) = \hat{G}(k) + \hat{H}(k), \forall k \in O(i) \text{ and } \hat{F}(j) \neq \hat{F}(k) \quad (36)$$

With Manhattan and misplaced distance concepts, we have

$$\hat{F}(j) \leq \hat{F}(k) - C_{\min} \quad (37)$$

Therefore, it can be derived that

$$\hat{F}_{DSA*}(j) = \hat{G}(j) + \hat{H}_{DSA*}(j) = \hat{G}(j) + \hat{H}(j) - \varepsilon C_{\min}/\exp(\hat{H}(j)) < \hat{F}(j) \quad (38)$$

$$\hat{F}_{DSA*}(k) = \hat{G}(k) + \hat{H}(k) - \varepsilon C_{\min}/\exp(\hat{H}(k)) \geq \hat{F}(k) - C_{\min}/\exp(\hat{H}(k)) > \hat{F}(k) - C_{\min} \quad (39)$$

$$\hat{F}_{DSA*}(j) < \hat{F}(j) \leq \hat{F}(k) - C_{\min} < \hat{F}_{DSA*}(k) \quad (40)$$

Eq. (40) ends the proof by ensuring that $\hat{H}_{DSA*}$ will still pick the same current node as $\hat{H}$ to further explore and does not prefer other existing nodes.

**Proposition 5**: The modified heuristic function improves the algorithm's efficiency by preventing it from jumping back to inner layer nodes with the same $F$ value.

**Proof**: When multiple shortest paths exist, and the heuristic function perfectly estimates the cost to the goal state, there will be multiple nodes in both inner and outer layers with the same smallest $F$ value. In this condition, the deterministic A* will recurrently explore inner layers nodes rather than expanding the path. A random factor generated with a fixed rule can break the equalities in the same layer but cannot necessarily prevent the algorithm from jumping back and forth between different layers. To this end, the heuristic function in Eq. (33) is designed to be hierarchical and ensures that the DSA* always prefers expanding the path in such conditions. The key is to make sure that the random part in Eq. (33) is monotonically decreasing, which is proved as follows.

$$\Psi = -\varepsilon C_{\min} / \exp(\hat{H}), \hat{F}(i) = \hat{F}(j), \ \hat{G}(i) < \hat{G}(j), \ \varepsilon \in (\exp(-C_{\min}), 1) \tag{41}$$

$$\Psi(i)/\Psi(j) = \exp(\hat{H}(j) - \hat{H}(i)) \, \varepsilon(i)/\varepsilon(j) \tag{42}$$

$$\Psi(i)/\Psi(j) = \exp(\hat{H}(j) - \hat{H}(i)) \, \varepsilon(i)/\varepsilon(j) < \exp(C_{\min}) \exp(\hat{H}(j) - \hat{H}(i)) \tag{43}$$

where $\Psi$ is the random part in Eq. (35); $i$ denotes a node in the inner layer; and $j$ denotes a node in the outer layer. With Manhattan and misplaced heuristic functions, we have

$$\hat{H}(j) - \hat{H}(i) \geq C_{\min} \tag{44}$$

Then, Eq. (43) can be derived as

$$\Psi(i)/\Psi(j) < \exp(C_{\min}) \exp(\hat{H}(j) - \hat{H}(i)) \leq \exp(C_{\min} - C_{\min}) \leq 1 \tag{45}$$

Therefore, we have Eq. (46) which concludes the proof.

$$\hat{F}_{DSA*}(j) = \hat{F}(j) + \Psi(j) < \hat{F}(i) + \Psi(i) = \hat{F}_{DSA*}(i) \tag{46}$$

Based on **Proposition 3** to **5**, it can be concluded that the merit of the proposed hierarchical stochastic heuristic function lies in: (1) maintaining optimality; (2) breaking uniform sub-graphs and returning other shortest paths with the same cost in different iterations; (3) having the same efficiency as the deterministic A* when the shortest path is unique; and (4) improving the efficiency of path expansion in the presence of multiple optimal paths. Note that the specific function of Eq. (34) should depend on the deterministic heuristic function, which could be easily achieved through **Proposition 3** to **5**.

## 6   CASE STUDY

A case study is provided to demonstrate the proposed algorithm for cooperative sorting. For simplification, it is assumed that all vehicles are running with the same cruising speed in all examples. In addition, the coefficients for the edge cost function is set up as $\beta_{long} = \beta_{lc} = 1$, with equal preference for longitudinal and lateral movements.

### 6.1 Result of the Deterministic A* Algorithm

The initial and final (goal) permutations of the CAV platoon in Fig. 3 are used in this example. The permutation of the CAVs is firstly discretized as a $4 \times 3$ grid system based on the method described in section 3.1. Since each vehicle has a unique destination, the Manhattan distance is used for calculating the heuristic function. The result of the deterministic A* algorithm is shown in Table 1, where the MaxID indicates the number of nodes that have been explored and ParentID is used for tracking the shortest path. It can be found that the goal node found is 13. The estimated distance to the goal node (the $F$ value) initially underestimated the real value in the first two steps but gradually approached the real distance without overestimation. The number of explored nodes is 652, which is only 0.09% of the total number of nodes in the graph domain, $A_{12}^6 = 665{,}280$. Fig. 12 shows another perspective of the A* algorithm with all explored nodes, where the grey dots are the open list nodes; the blue dots are explored current nodes during the path-finding process; and the red ones indicate the nodes that belong to the shortest path.

The concentric circles in Fig. 12 are arranged by the $G$ value. As the path approaches the goal node, the number of explored nodes generally decreases because the shortest path becomes clearer in the later stages of the algorithm implementation. Furthermore, the algorithm may jump back and forth between different layers judging by many current nodes with the same $F$ values in various layers, suggesting that the algorithm could jump back and forth between different layers, which undermines the efficiency of the algorithm.

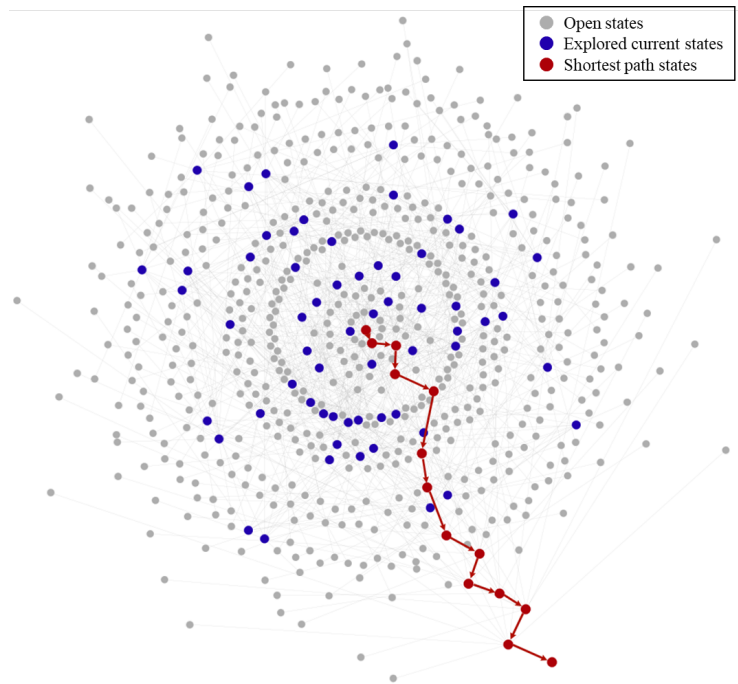

**Fig. 12.** The expanded subgraphs explored by the A* algorithm

Table 1. The result of the deterministic A* algorithm

| | Initial | Step-1 | Step-2 | Step-3 | Step-4 | Step-5 | Step-6 | Step-7 | Step-8 | Step-9 | Step-10 | Step-11 | Step-12 | Step-13 |
|---|---|---|---|---|---|---|---|---|---|---|---|---|---|---|
| Position 1 | 0 | 0 | 0 | 0 | 0 | 0 | 0 | 0 | 0 | 0 | 0 | 0 | 0 | 0 |
| Position 2 | 0 | 0 | 0 | 0 | 0 | 0 | 0 | 0 | 0 | 0 | 0 | 0 | 0 | 0 |
| Position 3 | 0 | 0 | 0 | 0 | 0 | 0 | 0 | 0 | 0 | 0 | 0 | 0 | 0 | 0 |
| Position 4 | C | C | C | C | C | C | 0 | A | A | A | A | A | A | A |
| Position 5 | F | 0 | D | D | D | 0 | C | C | C | C | C | 0 | B | B |
| Position 6 | D | D | 0 | 0 | 0 | 0 | 0 | 0 | 0 | 0 | C | C | C | C |
| Position 7 | A | A | A | A | A | A | A | 0 | D | D | D | D | D | D |
| Position 8 | 0 | F | F | F | 0 | D | D | D | 0 | B | B | B | 0 | E |
| Position 9 | E | E | E | 0 | F | F | F | F | F | F | F | F | F | F |
| Position 10 | 0 | 0 | 0 | 0 | 0 | 0 | 0 | 0 | 0 | 0 | 0 | 0 | 0 | 0 |
| Position 11 | B | B | B | B | B | B | B | B | B | 0 | E | E | E | 0 |
| Position 12 | 0 | 0 | 0 | E | E | E | E | E | E | E | 0 | 0 | 0 | 0 |
| G Value | 0 | 1 | 2 | 3 | 4 | 5 | 6 | 7 | 8 | 9 | 10 | 11 | 12 | 13 |
| H Value | 11 | 10 | 9 | 10 | 9 | 8 | 7 | 6 | 5 | 4 | 3 | 2 | 1 | 0 |
| F Value | 11 | 11 | 11 | 13 | 13 | 13 | 13 | 13 | 13 | 13 | 13 | 13 | 13 | 13 |
| ID | 0 | 3 | 61 | 73 | 560 | 566 | 588 | 597 | 605 | 617 | 624 | 626 | 638 | 652 |
| MaxID | 0 | 58 | 67 | 555 | 562 | 587 | 593 | 600 | 607 | 617 | 624 | 632 | 641 | 652 |
| ParentID | 0 | 0 | 3 | 61 | 73 | 560 | 566 | 588 | 597 | 605 | 617 | 624 | 626 | 638 |

## 6.2 Results of the Sorting Time Optimization

The resulting vehicle trajectories are shown in Fig. 13. It is found that 13 movement steps are required to complete the sorting process by implementing the proposed methods in Section 3. When the multi-movement principle is adopted at the conservative level (Eq. (22) to Eq. (27)), the shortest path can be found in only 9 movement steps to achieve the goal permutation, with multiple movements in step 1, 2 and 5; see Fig. 14. At the aggressive level, only 4 steps are needed to achieve the goal permutation, as shown in Fig. 15. Here there are multiple movements in every step.

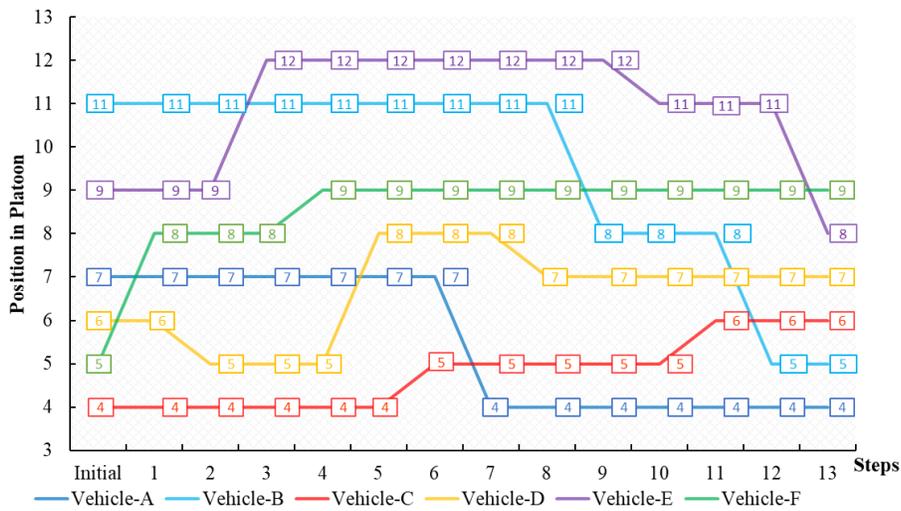

**Fig. 13.** The trajectory of the shortest path with stepwise-principle

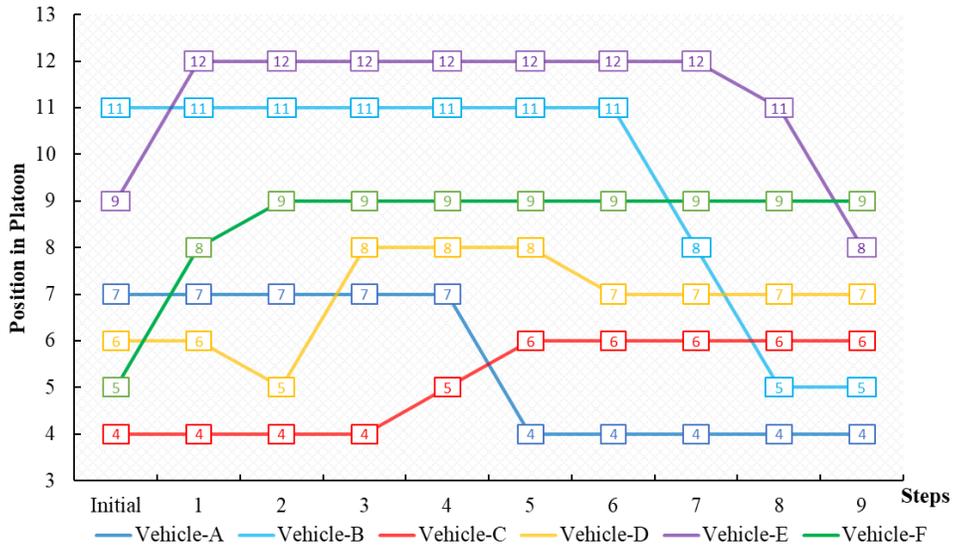

**Fig. 14.** The optimized trajectory in the conservative simultaneous level

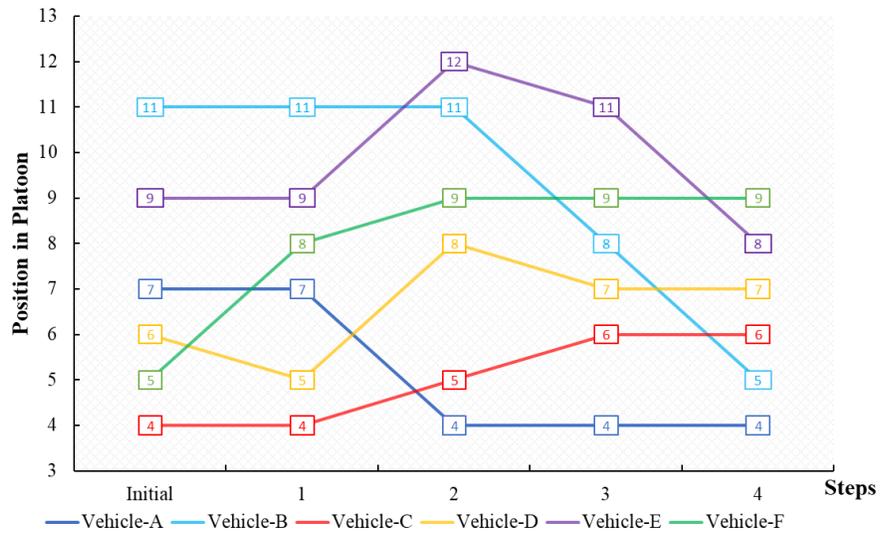

**Fig. 15.** The optimized trajectory in aggressive simultaneous level

## 6.3 Results of the DSA* Algorithm

To illustrate the benefit of the DSA* algorithm, we pick two goal states, as shown in Table 2, and randomly generate 30 initial states. Assuming that vehicle A, B, and C are left-turning vehicles and D,

E and F are through vehicles, the goal states correspond to two possible ways of implementing the tandem design. The experiment is conducted with a 2.6 GHz Intel I7 CPU and the R language. Four experiments are conducted: (1) deterministic A*, with all 30 initial samples and goal state 1; (2) DSA* with 8 representative initial samples and goal state 1; (3) DSA* with the same 8 initial samples and both goal state 1 and goal state 2; (4) DSA* and ILP algorithm (for multi-movement principle) with sample 27 and goal state 1 for 100 iterations to demonstrate solution convergence.

The result of experiment 1 is shown in Fig. 16, in terms of running time vs. the length of the shortest path (the $F$ value). The figure shows that the length of the path, representing the path complexity, significantly influences the solving time of the algorithm, such that the running time tends to increase exponentially with the path length. Nonetheless, the performance of the deterministic A* is stochastic and thus, not practical enough. (Note that the result in Fig. 16 shows the performance trend, and the absolute running time varies with different computing servers and codes.)

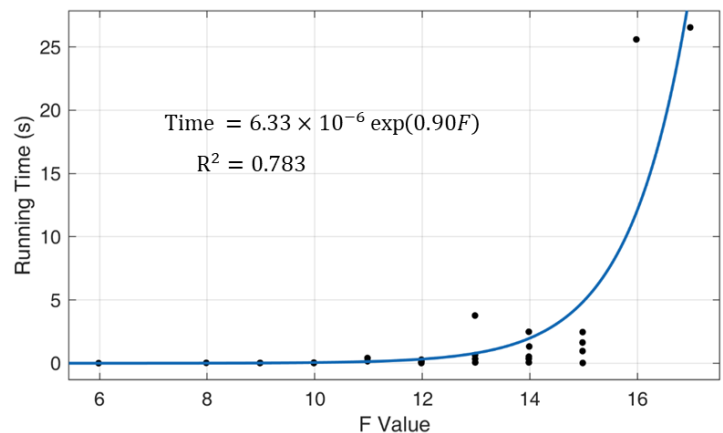

**Fig. 16.** The running time of the deterministic A* between 30 initial samples and goal state 1

**Table 2. Sample list**

| Sample ID | Position | | | | | | | | | | | |
|---|---|---|---|---|---|---|---|---|---|---|---|---|
| | 1 | 2 | 3 | 4 | 5 | 6 | 7 | 8 | 9 | 10 | 11 | 12 |
| Initial-1 | C | 0 | 0 | 0 | E | 0 | D | A | 0 | F | 0 | B |
| Initial-2 | D | 0 | C | 0 | F | A | 0 | B | 0 | 0 | 0 | E |
| Initial-3 | C | 0 | B | 0 | A | 0 | 0 | E | 0 | F | D | 0 |
| Initial-4 | C | B | E | 0 | A | D | 0 | 0 | 0 | 0 | 0 | F |
| Initial-5 | 0 | C | 0 | 0 | E | A | 0 | F | 0 | B | D | 0 |
| Initial-6 | B | 0 | C | 0 | F | 0 | 0 | 0 | E | D | A | 0 |
| Initial-7 | F | E | A | 0 | 0 | 0 | 0 | B | C | 0 | D | 0 |
| Initial-8 | B | E | 0 | F | 0 | 0 | C | 0 | 0 | D | 0 | A |
| Initial-9 | 0 | B | 0 | A | 0 | 0 | E | C | D | 0 | F | 0 |
| Initial-10 | 0 | A | 0 | E | 0 | 0 | 0 | D | B | F | C | 0 |
| Initial-11 | 0 | 0 | F | 0 | A | C | 0 | 0 | D | E | 0 | B |
| Initial-12 | D | 0 | C | E | 0 | 0 | B | 0 | F | 0 | 0 | A |
| Initial-13 | D | 0 | A | B | E | F | 0 | 0 | 0 | C | 0 | 0 |
| Initial-14 | F | D | 0 | 0 | 0 | C | 0 | A | E | 0 | B | 0 |
| Initial-15 | 0 | 0 | 0 | A | D | 0 | C | B | F | 0 | 0 | E |
| Initial-16 | B | F | A | C | D | E | 0 | 0 | 0 | 0 | 0 | 0 |
| Initial-17 | 0 | 0 | C | E | 0 | F | 0 | A | B | D | 0 | 0 |
| Initial-18 | 0 | E | 0 | 0 | B | C | 0 | D | 0 | A | F | 0 |
| Initial-19 | A | D | C | 0 | 0 | 0 | E | F | 0 | 0 | B | 0 |
| Initial-20 | B | 0 | D | 0 | A | 0 | 0 | F | C | 0 | E | 0 |
| Initial-21 | A | D | 0 | 0 | B | 0 | F | 0 | C | 0 | 0 | E |
| Initial-22 | B | 0 | 0 | A | C | F | 0 | D | 0 | 0 | E | 0 |
| Initial-23 | D | 0 | F | 0 | 0 | 0 | A | 0 | B | E | C | 0 |
| Initial-24 | C | F | A | D | 0 | B | E | 0 | 0 | 0 | 0 | 0 |
| Initial-25 | 0 | 0 | 0 | E | C | D | A | 0 | 0 | F | 0 | B |
| Initial-26 | E | B | 0 | 0 | 0 | F | A | 0 | C | 0 | 0 | D |
| Initial-27 | 0 | D | 0 | B | E | 0 | 0 | 0 | 0 | C | F | A |
| Initial-28 | A | B | F | 0 | 0 | C | 0 | 0 | E | D | 0 | 0 |
| Initial-29 | B | D | F | 0 | 0 | 0 | 0 | E | A | 0 | C | 0 |
| Initial-30 | E | 0 | 0 | D | C | A | 0 | 0 | 0 | B | F | 0 |
| Goal-1 | 0 | 0 | 0 | A | B | C | D | E | F | 0 | 0 | 0 |
| Goal-2 | 0 | 0 | 0 | D | E | F | A | B | C | 0 | 0 | 0 |

In experiment 2, 8 samples with different $F$ values were randomly selected for a more detailed investigation. This experiment was conducted without any time limit setup. The distributed system was simulated with 10 threads of the CPU, with each thread processing 3 runs of the DSA*. The running time performance is shown in Table 3. The result shows that the running time is significantly lower compared to the deterministic A*. However, for unfavorable conditions such as sample 30, the running time is still undesirable in practice. The main reason for this poor performance is that the initial state in sample 30 is not favorable to be sorted to goal state 1, indicating that goal state 1 might just be a poor choice as a goal state.

In experiment 3, both goal states are considered in the DSA* algorithm without any time limit setup. The results in Table 3 indicate that DSA* jointly with multiple goal states provide a more efficient and consistent solution. The average running time for finding an optimal solution for the tandem concept is 0.192s for the representative samples.

Table 3. The running time of DSA* against deterministic A*

| ID | F-value | Experiment 1(s) Deterministic A* | Experiment 2 (s) DSA* with goal state 1 | | | Experiment 3 (s) DSA* with both goal states | | |
|----|---------|----------------------------------|------|------|------|------|------|------|
| | | | Min | Max | Mean | Min | Max | Mean |
| 22 | 6 | 0.008 | 0.008 | 0.030 | 0.009 | 0.013 | 0.047 | 0.016 |
| 28 | 8 | 0.032 | 0.031 | 0.067 | 0.050 | 0.053 | 0.105 | 0.081 |
| 9 | 10 | 0.053 | 0.044 | 0.051 | 0.047 | 0.070 | 0.093 | 0.075 |
| 11 | 12 | 0.126 | 0.124 | 0.234 | 0.186 | 0.200 | 0.395 | 0.296 |
| 14 | 14 | 0.558 | 0.474 | 0.740 | 0.568 | 0.024 | 0.064 | 0.036 |
| 29 | 15 | 2.600 | 0.545 | 0.734 | 0.649 | 0.616 | 0.744 | 0.660 |
| 27 | 17 | 27.216 | 2.583 | 3.250 | 2.780 | 0.028 | 0.616 | 0.221 |
| 30 | 16 | 25.480 | 15.235 | 16.201 | 15.570 | 0.027 | 0.489 | 0.154 |
| Average running time (s) | | 7.009 | | 2.483 | | | 0.192 | |

In experiment 4, the objective is to demonstrate the improvement of DSA* by allowing multiple movements simultaneously. For this, sample 27, which has the most complicated shortest path in our sample set, is used for the experiment. The DSA* algorithm is run for 100 iterations to simulate a distribution system with 100 threads. For each returned trajectory, the ILP optimization is conducted based on the conservative and aggressive principles. Fig. 17(a) shows the results of the optimization. It verifies our premise that the equal shortest paths could lead to different ILP optimization results. For the aggressive level of multiple simultaneous movements, the number of steps to finish sorting is concentrated around 5 to 8 steps, involving 17 total vehicle movements. For the conservative level, the number ranges from 8 to 16 steps. The aggressive group has more concentrated results primarily due to the relaxed constraint in Eq. (33) compared to Eq. (27), so that various paths can lead to similar optimization results.

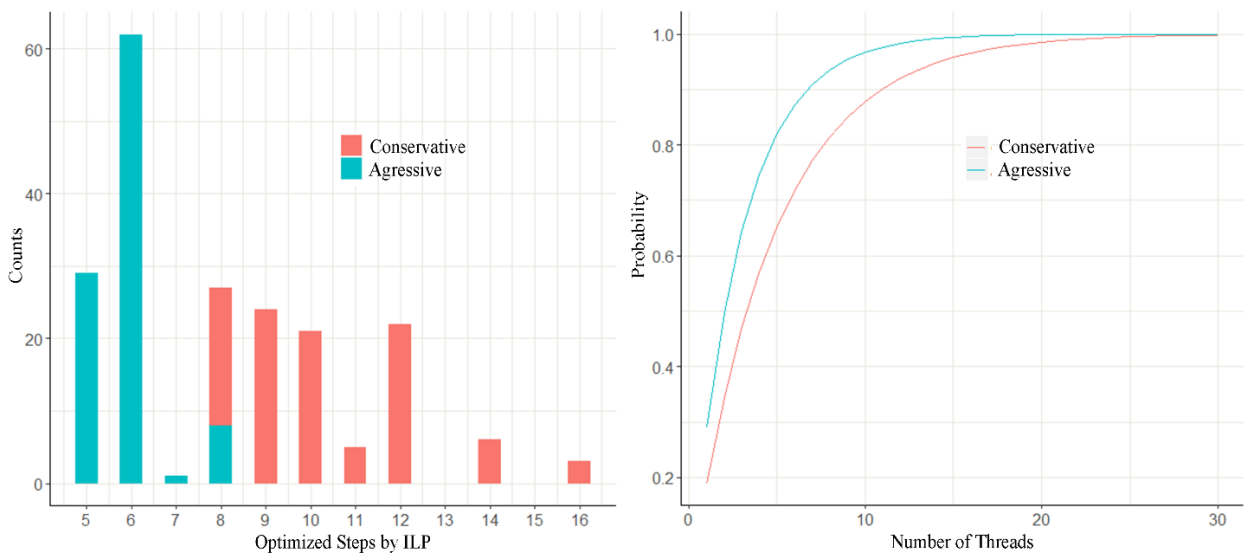

**Fig. 17.** The DSA* results of sample 27 with 100 iterations

The probability of getting the best optimization result (i.e., the lowest number of steps) vs. the number of threads is shown in Fig. 17(b). Here we are assuming that experiment 4 has resulted in global optimum solutions, 8 and 5 steps respectively for the conservative and aggressive level multi-movement principles. It shows that the probability grows exponentially with respect to the number of threads, and 30 threads can ensure the best result with 99.99% confidence.

# 7  CONCLUSION AND DISCUSSION

This study developed a method for sorting connected and automated vehicles to attain a desirable permutation of vehicles in a platoon. In the proposed method, the CAV platoon was firstly discretized into a grid system and then a mathematical matrix. Then the cooperative platoon sorting problem was modeled as a path-finding problem in the graphic domain, where the nodes represent permutations of vehicle platoons, and the shortest path indicates the optimal solution of the sorting problem. The deterministic A* algorithm was used to find the shortest path in the graphic space. The method was further refined to allow multiple simultaneous vehicle movements to improve algorithm efficiency, which can be solved by an integer linear programming (ILP) method with a coordinate system designed in this study. To address the issues of algorithm running time and handling multiple shortest paths, a distributed stochastic A* algorithm (DSA*) was developed by introducing random disturbances with a hierarchical structure to the edge costs to break uniform paths (with equal path cost). Numerical experiments demonstrated effectiveness of the proposed DSA* method with shorter sorting time and significantly improved algorithm running time in terms of average value and consistency across different initial states. In addition, we found that the optimization performance can be further improved by increasing the number of processes in the distributed computing system.

Several limitations of this study are notable. The proposed method relies on 100% penetration of CAVs. Future studies are desired to investigate the relationship between the penetration rate of CAVs and the performance of the proposed framework, and to develop a sorting strategy for platoons mixed with non-cooperative vehicles. The control framework developed in this study can serve as a building block for these extensions. Another limitation is the assumption of well-defined vehicle platoons. In reality, vehicles might be distributed more randomly on a corridor, rather than forming clear platoons. In this situation, the traffic needs to be divided into several "platoons" to apply the proposed algorithms. This process will depend on several factors, such as the range of V2V communication and desired states at downstream facilities. This topic is beyond the scope of the present paper and needs future investigations.

**APPENDIX A**

In this appendix, an example is provided to explain the A* algorithm. Assume that a tourist intends to drive from the current position *I* city to his destination *T* city. There are 8 cities indexed with *A* to *H* connecting the origin to destination, as shown in Fig. 18. In the figure, the numbers on the lines denote the cost of driving between cities. Background information is provided that the cost from any city to *T* city is no larger than the flight cost in Table 4. The problem is to find the shortest path for the tourist to drive from *I* city to *T* city.

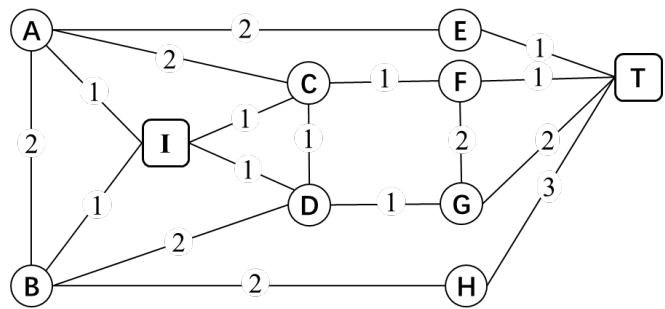

**Fig. 18.** The city map and driving cost

**Table 4.** The cost of flight to *T* city

| Nodes | A | B | C | D | E | F | G | H |
|---|---|---|---|---|---|---|---|---|
| Flight cost to T | 3 | 3 | 2 | 2.5 | 1 | 1 | 1.5 | 2 |

Based on the procedure of A* algorithm, we first put the origin point *I* city into the open list. Since there is only one city on the open list, I city would be chosen as the current list in step 2. In the meanwhile, the adjacent cities *A, B, C,* and *D* are added into the open list and the *F* values can be calculated with Eq. 1, as shown in Fig. 19(a). Note that the cost of flight here is used as the heuristic distance. In the open list, C city has the minimum estimated distance and, therefore, is chosen as the current node in the next iteration in Fig. 19(b). Notably, in Fig. 19(b), city *A* and *D* are common children nodes of the city *I* and *C*. In this condition, we should check which parent node provides a better path with respect to travel cost. The *F* values of *A* and *D* are then calculated assuming that *I* and *C* are parent nodes, respectively. For city *A*, as an example, the *F* value from *C* city is $F = 1 + 2 + 3 = 6$ but from *I* city $F = 1 + 3 = 4$. Obviously, *I* city is a better parent node for *A* city and, therefore, the *F* value of open node A is recorded as 4. This checking process should always be conducted when common parent nodes exist. In Fig. 19(c), the *F* city is chosen as the current node since it has the minimum *F* value. The algorithm terminates in Fig. 19(d) by finding the destination node *T*.

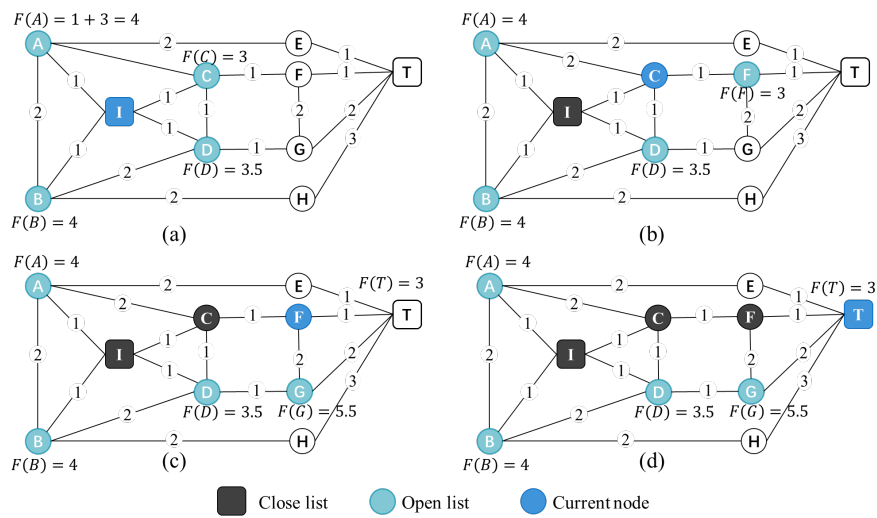

**Fig. 19.** The tourist example of A* algorithm

In the tourist example, it could be found that the heuristic distance is gradually replaced by the real distance as the path approaches the destination. The heuristic function provides direction to the destination so that the algorithm only examines necessary nodes. For instance, E city and H city were not even considered in Fig. 19.

## ACKNOWLEDGEMENT

This study was jointly sponsored by the National Science Foundation through Award CMMI 1536599 and National Key R&D Program of China (2018YFB1600900).